\documentclass[aps,amsmath,twocolumn,showpacs]{revtex4}
\usepackage{graphicx}
\usepackage{epstopdf}
\begin{document}
\title{Tunneling induced giant Photonic Spin Hall effect in quantum wells}

\author{Fazal Badshah$^{1,2}$}
\author{Imed Boukhris$^3$}
\author{Mohamed Sultan Al-Buriah$^4$}
\author{Yuan Zhou$^{1}$}
\author{Muhammad Idrees$^{5}$}
\author{Ziauddin$^6$}
\email{ziauddin@comsats.edu.pk}

\address{$^1$ Shiyan Key Laboratory of Electromagnetic Induction and Energy Saving Technology, School of Artificial Intelligence, Hubei University of Automotive Technology, Shiyan 442002, People’s Republic of China}
\address{$^2$ Applied Science Research Center,  Applied Science Private University, Amman 11931, Jordan}
\address{$^3$ Department of Physics, Faculty of Science, King Khalid University, P.O. Box 960, Abha, Saudi Arabia} 
\address{$^4$ Department of Physics, Sakarya University, Sakarya, Turkey} 
\address{$^{5}$ School of Physics, Zhejiang Normal University, Jinhua 321004, People's Republic of China}
\address{$^6$ Quantum Optics Lab. Department of Physics, COMSATS University Islamabad, Pakistan}

\begin{abstract}
 We propose a theoretical investigation of the photonic spin Hall effect (PSHE) in a mid-infrared probe field by employing an asymmetric double AlGaAs/GaAs quantum well as the intracavity medium. The system is designed such that an external control beam together with tunable tunneling barriers regulates the quantum interference of the probe tunneling process. This configuration enables precise manipulation of the PSHE for both horizontally and vertically polarized components of light. Our analysis reveals the emergence of a giant horizontal PSHE in the quantum well–based cavity system. Moreover, by incorporating absorptive and gain-assisted cavity slabs, the horizontal PSHE is further amplified, leading to an even more pronounced photonic spin separation. The results provide novel insights into light–matter interactions in semiconductor quantum wells and suggest an effective route for enhancing and controlling the PSHE in mid-infrared photonic devices.

\end{abstract}
\date{\today}

\maketitle

\section{Introduction}
The spin Hall effect originally arises in solid-state systems, where the transverse shift of particles such as electrons occurs due to spin–orbit coupling \cite{J-Sinova-rmp, B-A-Bernevig-prl}. The spin-orbit interaction establishes in these systems due to the distinct forces acting on spin-down and spin-up states propagating in a potential \cite{T-Jungwirth-n}. This leads to a transverse particle displacement, a phenomenon known as the spin Hall effect (SHE) \cite{J-E-Hirsch-prl}. The spin Hall effect has been investigated in a wide range of physical systems. Notable examples include two-dimensional materials \cite{C-Safeer-n}, graphene \cite{C-L-Kane-prl}, semiconductors \cite{Y-K-Kato-sc}, and topological insulators \cite{B-A-Bernevig-prl}.
The photonic counterpart to this electronic phenomenon is the photonic spin Hall effect (PSHE), which describes a transverse spin-dependent shift of photons interacting with a material interface. The PSHE establishes from the distinct behavior of right-circularly and left-circularly polarized light, directly analogous to how spin-up and spin-down electrons experience different forces. The role of the electronic potential is played by the refractive index gradient of the medium, leading to a similar spin-dependent splitting for photons. Upon interacting with the interface of a coherent medium, spin-orbit interaction causes left-circularly and right-circularly polarized photons to undergo opposite transverse shifts, perpendicular to the plane of incidence, thus constituting the PSHE \cite{M-Kim-pr}. 
In recent years, researchers have devoted significant attention to the PSHE as a means to control the spin-dependent characteristics of photons in various optical media. These media include surface plasmon resonance schemes \cite{L-Salasnich-pra, X-J-Tan-ol, R-G-Wan-pra}, semiconductors \cite{J-M-Mnard-ol}, metamaterials \cite{X-Yin-sc}, and quantum materials \cite{W-Kort-Kamp-prl}. 

Although the PSHE has been investigated in numerous atomic systems \cite{R-G-Wan-pra, J-Wu-op, M-Waseem-pra, M-Shah-arc}, none have demonstrated a substantial enhancement of the effect in the reflected probe field. This challenge was recently overcome by a tripod-type atomic configuration \cite{M-Abbas-pra}, which achieved a notable enhancement, underscoring the importance of multi-level atomic coherence and field coupling for amplifying spin-dependent photonic shifts. 

Furthermore, quantum well systems present a range of advantages over atomic systems, particularly in terms of scalability, tunability, environmental compatibility, and fabrication. Unlike atomic systems, which require complex trapping and cooling mechanisms, quantum wells can be readily fabricated using established semiconductor techniques, enabling large-scale integration into modern electronic and photonic platforms. A key feature of quantum wells is their high degree of tunability; parameters such as bandgap and energy levels can be precisely engineered by varying the well’s width, depth, and material composition. Additionally, operating in solid-state environments offers enhanced control over external conditions, including temperature regulation, electric field gating, and magnetic field application. Notably, quantum well structures exhibit strong light–matter interactions, making them highly effective in optoelectronic devices such as quantum well lasers, detectors, and modulators.

The investigation of the PSHE has attracted significant attention due to its fundamental role in light–matter interactions and its promising applications in next-generation photonic and optoelectronic devices. A critical challenge in this domain lies in achieving large, tunable transverse spin-dependent shifts while minimizing optical losses. Addressing this challenge is essential for advancing the practical implementation of PSHE-based technologies. Motivated by the unique advantages offered by quantum well systems such as their high tunability, strong light–matter interaction, and compatibility with integrated photonic platforms this work proposes a novel quantum well structure designed to control and enhance the PSHE for both horizontally and vertically polarized light. By leveraging the engineered properties of the quantum well system, we aim to realize enhanced and controllable transverse spin shifts, paving the way for more efficient and functional photonic devices.

\textbf{Model}

\begin{figure}
\centering
\includegraphics[width=3.5in]{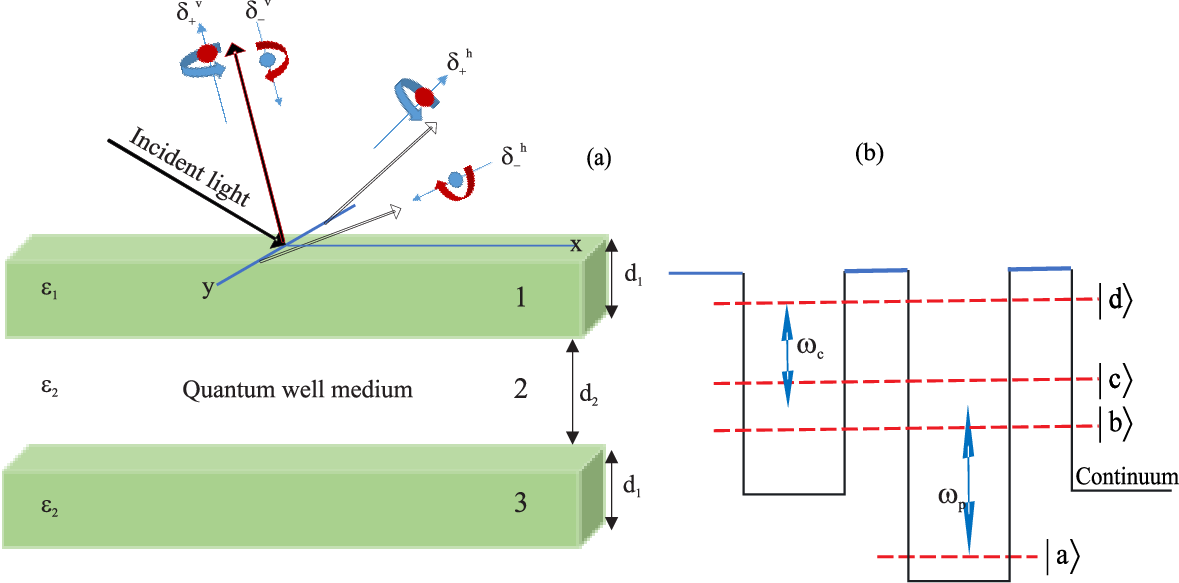}
\caption{(a) The schematics where a probe light interact with a cavity having gain-assisted three-level atoms (b) the energy level configuration of three-level atomic medium.}
\label{figure1}
\end{figure}

We suggest a TM-polarized probe field incident on a cavity at an angle $\theta$ with respect to the interface. The cavity consists of a three-layer configuration, where the outer slabs (slab 1 and slab 3) serve as cavity walls, and semiconductor quantum wells having the configuration of four sub-bands is embedded with in the central intracavity region (slab 2) as illustrated in Fig. \ref{figure1}(a). The permittivities of the boundary slabs are denoted by $\epsilon_1$ and $\epsilon_3$, respectively, each with thickness $d_1$, while the intermediate atomic medium has thickness $d_2$. In this arrangement, $\epsilon_1$ and $\epsilon_3$ remain fixed, whereas the effective permittivity of the intracavity medium can be dynamically tuned according to the relation \cite{Ziauddin-pra-g}
\begin{eqnarray}
\epsilon_2 = 1 + \chi,
\end{eqnarray}
where $\chi$ represents the susceptibility of the intracavity medium. The susceptibility plays a pivotal role in governing both the dispersive and absorptive (gain or loss) properties of the system. Importantly, $\chi$ can be coherently controlled by the external driving laser fields \cite{M-O-Scully-book, Freedhoff-pra}, which makes the cavity highly versatile in tailoring light–matter interactions. Within this framework, our next step is to derive and analyze the susceptibility of the proposed intracavity medium, which enables us to explore novel regimes of dispersion engineering and polarization-dependent PSHE.

Figure 1(b) illustrates the asymmetric quantum well structure (serving as the intracavity medium), along with the corresponding conduction band profile and wave function. This structure can be fabricated sequentially from left to right. A thick $Al_{0.4} Ga_{0.6}$As  barrier is followed by a 6.8 nm-wide shallow quantum well composed of $Al_{0.16} Ga_{0.84}$As. This shallow well is separated from a 7.7 nm-wide GaAs deep quantum well on the right by a 3.0 nm-wide $Al_{0.4} Ga_{0.6}$As  potential barrier. Following this, a thin (1.5 nm) $Al_{0.4}Ga_{0.6}$As barrier separates the deep well from the adjacent thick layer of $Al_{0.16}Ga_{0.84}$As. The energies are $E_a = 46.7$ meV for the deep well's ground state $|a\rangle$ and $E_d = 296.3$ meV for the shallow well's first excited state $|d\rangle$. Two intermediate levels, $|b\rangle$ and $|c\rangle$ with energies $E_b=174.8$ meV and $E_c=13.5$ meV, are formed. These arise from the mixing of the shallow well's ground state and the deep well's first excited state via resonant tunneling. To simplify the model, we assume the coherent control beam ($\omega_c$) is far-detuned from the cavity resonances. This assumption ensures the beam interacts homogeneously with both the $|b\rangle \leftrightarrow |d\rangle$ and $|c\rangle \leftrightarrow |d\rangle$ transitions. Further, A weak probe beam ($\omega_p$) simultaneously excites the $|a\rangle \leftrightarrow |b\rangle$ and $|a\rangle \leftrightarrow |c\rangle$ transitions.

Within the rotating-wave and dipole approximations, the equations of motion become as
\begin{eqnarray}
\dot{B}_1&=&i g \Omega_p B_2+i \Omega_pB_3,\nonumber\\
\dot{B_2}&=&i(i\gamma_2-\Delta_p+\Delta)B_2+ig\Omega_pB_1+if\Omega_cB_4+\alpha B_3,\nonumber\\
\dot{B_3}&=&i(i\gamma_3-\Delta_p+\Delta)B_3+i\Omega_pB_1+i\Omega_cB_4+\alpha B_2,\nonumber\\
\dot{B_4}&=&i(i\gamma_4-\Delta_p+\Delta_c)B_4+if\Omega_cB_2+i\Omega_cB_3,\nonumber\\\label{d-matrix}
\end{eqnarray} 
where $\Delta_p=(\frac{E_a+E_c}{2}-E_a)-\omega_p$, $\Delta_c=(E_d-\frac{E_a+E_c}{2})-\omega_c$ where $2\Delta=E_c-E_b$ is the energy splitting between $|b\rangle$ and $|c \rangle$. Here, $\Omega_p$ and $\Omega_c$ are the Rabi frequencies of the probe and control fields, respectively. Also, the parameters $g = \mu_{ba} / \mu_{ca}$ and $f = \mu_{bd} / \mu_{cd}$ represent the ratios between the transition dipole moments for the specified energy levels. We define the total decay rate $\gamma_i$ for each state $i$ ($i = a, b, c, d$). This total rate consists of a population decay component $\gamma_{il}$ and a dipole dephasing component $\gamma_{id}$. Finally, the cross-coupling between states $|b\rangle$ and $|c\rangle$, which arises from tunneling into the electronic continuum, is quantified by the parameter $\alpha = \sqrt{\gamma_{bl} \gamma_{cl}}$ \cite{J-Faist-np, H-Schmidt-apl}. The Fano-type interference induced by the strong tunneling is characterized by the parameter $p = \alpha / \sqrt{\gamma_b \gamma_c}$, which has a value of 0.83. Here, $p$ quantifies the interference strength, where $p=0$ signifies no interference and $p=1$ signifies strong interference.  

Using the steady state solution of Eq. (\ref{d-matrix}), we can find the susceptibility as 
\begin{eqnarray}
\chi=\frac{i\beta(A_1\gamma_4+(f-g)^2\Omega_c^2)}{A_2\gamma_4+A_3\Omega_c^2},	\label{susc}
\end{eqnarray}	
where $A_1=-i\Delta+ig^2\Delta+2 g \alpha+\gamma_2+g^2\gamma_3$, $A_2=\Delta^2-\alpha^2-i\Delta \gamma_3+\gamma_2(i\Delta+\gamma_3)$, $A_3=-i\Delta+if^2\Delta+2f\alpha+\gamma_2+f^2\gamma_3$ and $\beta=N\mu_{ca}^2/\epsilon_0\hbar$.

\subsection{Theoretical formalism of the PSHE}

Once the probe beam propagates through the cavity, it can either be transmitted through or reflected back from the cavity interfaces. We consider the probe beam with both TE and TM polarizations incident at an angle $\theta$, as illustrated in Fig. \ref{figure1}(a). Upon reflection, the probe field can be decomposed into right- and left-circular polarization components. These circular components interact differently with the cavity system and experience distinct spin-dependent phase variations. As a consequence, they undergo opposite transverse displacements in both horizontal and vertical directions. This spin-dependent deflection gives rise to the PSHE, enabling a direct manifestation of spin–orbit interaction of light within the proposed cavity configuration.
To establish the mathematical framework of the PSHE, we begin by evaluating the reflection coefficients associated with the TE- and TM-polarized components of the probe field within the cavity. For this purpose, we employ the transfer matrix method, which allows the reflection coefficient of the TE-polarized component to be written as
\begin{equation}
P_{i}=
\begin{pmatrix}
\text{cos}(k_{x}^i d_i) & \frac{i\text{sin}(k_{x}^i d_i)}{q_{i}} \\iq_{i}\text{sin}(k_{x}^i d_i) & \text{cos}(k_{x}^i d_i) \label{tra-mat}
\end{pmatrix}.
\end{equation}
Here, $k_{x}^i=\sqrt{\epsilon_i k^2-k_z^2}$ represents the $x$-component of the wave vector in the $i$-th layer, while $d_j$ denotes the thickness of the $j$-th layer for $i=1,2,3$. In addition, we define $q_i = \tfrac{k_{x}^i}{k}$. For the three-layer cavity under consideration, the overall transfer matrix can then be written as
\begin{equation}
P=P_{1}. P_{2}. P_{3}.
\end{equation}
The expression for the reflection coefficient ($r_e$) is derived from the solution to the governing equations as
\begin{equation}
r_e=\frac{q_0(P_{22}-P_{11})-(q_0^2P_{12}-P_{21})}{q_0(P_{22}+P_{11})-(q_0^2P_{12}+P_{21})},\label{refe}
\end{equation}
where $q_{0}=\frac{k_{x}}{k}$.
Further, the transfer matrix for the i-th layer under TM polarization is defined as
\begin{equation}
U_{i}=
\begin{pmatrix}
\text{cos}(k_{x}^i d_i) & \frac{i\text{sin}(k_{x}^i d_i)}{p_{i}} \\ip_{i}\text{sin}(k_{x}^i d_i) & \text{cos}(k_{x}^i d_i) \label{tra-mat2}
\end{pmatrix},
\end{equation}
where $k_{x}^{(i)} = \sqrt{\epsilon_i k^2 - k_z^2} / \epsilon_i$ and $p_{i} = q_{j} / \epsilon_i$. The total transfer matrix is then given by the product as 
\begin{equation}
U=U_{1}. U_{2}. U_{3}.
\end{equation}
The reflection coefficient for TM polarization is given by
\begin{equation}
r_m=\frac{q_0(U_{22}-U_{11})-(q_0^2U_{12}-U_{21})}{q_0(U_{22}+U_{11})-(q_0^2U_{12}+U_{21})}.\label{refm}
\end{equation}
The reflection coefficients $r_e$ and $r_m$ (Eqs. \ref{refe} and \ref{refm}) are functions of both the cavity walls' static permittivity $\epsilon_1$ and their dynamic optical response. The expression $\epsilon_2=1+\chi$ for the effective permittivity provides a direct link between the atomic susceptibility and the system's reflection properties, thereby connecting the medium's response to the observable photonic output. An incident light beam exhibits the PSHE by undergoing a spin-dependent splitting into distinct right- and left-circularly polarized components. We study this effect by analyzing the incident and reflected electric fields. The angular spectrum for each photon spin state can be written as
\begin{eqnarray}
E_i^h=(E_{i+}+E_{i-})/\sqrt{2},
\end{eqnarray}
and 
\begin{eqnarray}
E_i^v=i(E_{i-}-E_{i+})/\sqrt{2},
\end{eqnarray}
where, $h$ and $v$ denote the horizontal and vertical polarization states, respectively.
Here, $(+)$ and $(-)$ indicate the right-circularly polarized and left-circularly polarized components, respectively. The reflection of a monochromatic Gaussian beam at the cavity is described by formulating the beam as a localized wave packet with a sufficiently narrow angular spectrum as
\begin{eqnarray}
E_i=\frac{\omega_0}{\sqrt{2\pi}}e^{-\frac{\omega_0^2(k_x2+k_y^2)}{4}},\label{ang-spec}
\end{eqnarray} 
with $\omega_0$ being the beam waist. The reflected field, $\mathbf{E}_r$, is expressed in terms of the incident field, $\mathbf{E}_i$, as follows as 
\begin{equation}
\begin{pmatrix}
E_r^h \\E_r^v 
\end{pmatrix}=
\begin{pmatrix}
r_m & \frac{k_{ry}\text{cot}(\theta)(r_m+r_e)}{k_0} \\-\frac{k_{ry}\text{cot}(\theta)(r_m+r_e)}{k_0} & r_e .\label{ref-mat}
\end{pmatrix}
\begin{pmatrix}
E_i^h \\E_i^v 
\end{pmatrix}
\end{equation}
Therefore, the reflected angular spectrum, obtained from Eqs. (\ref{ang-spec}) and (\ref{ref-mat}), is given by
\begin{eqnarray}
E_r^h=\frac{r_m}{\sqrt{2}}(e^{ik_{ry}\delta_{r+}^h}E_{r+}+e^{-ik_{ry}\delta_{r-}^h}E_{r-}),
\end{eqnarray}
and 

\begin{eqnarray}
E_r^v=i\frac{r_e}{\sqrt{2}}(-e^{ik_{ry}\delta_{r+}^v}E_{r+}+e^{-ik_{ry}\delta_{r-}^v}E_{r-}).
\end{eqnarray}

The PSHE in reflection arises from the spin-orbit interaction term $e^{\pm ik_{ry}\delta_r^{h,v}}$ \cite{H-Yu-na}. The magnitude of the effect for $h$ and $v$ polarizations is finally expressed as

\begin{eqnarray}
\delta_{\pm}^h=\mp \frac{\lambda}{2\pi}(1+|r_e|/|r_m|\times \text{cos}(\phi_e-\phi_m))\text{cot}\theta.\label{h-pshe}
\end{eqnarray}
\begin{eqnarray}
\delta_{\pm}^v=\mp \frac{\lambda}{2\pi}(1+|r_m|/|r_e|\times \text{cos}(\phi_m-\phi_e))\text{cot}\theta.\label{v-pshe}
\end{eqnarray}
Here, $\delta_{r\pm}^h$ and $\delta_{r\pm}^v$ denote the transverse PSHE shifts corresponding to horizontally and vertically polarized light, respectively. The parameter $\theta$ represents the incident angle, while $\lambda$ is the wavelength of the probe beam. In addition, the reflection-induced phase shifts are characterized by $\phi_e$ for the TE-polarized component and $\phi_m$ for the TM-polarized component of the incident light. 

\begin{figure}
	\centering
	\includegraphics[width=3.5in]{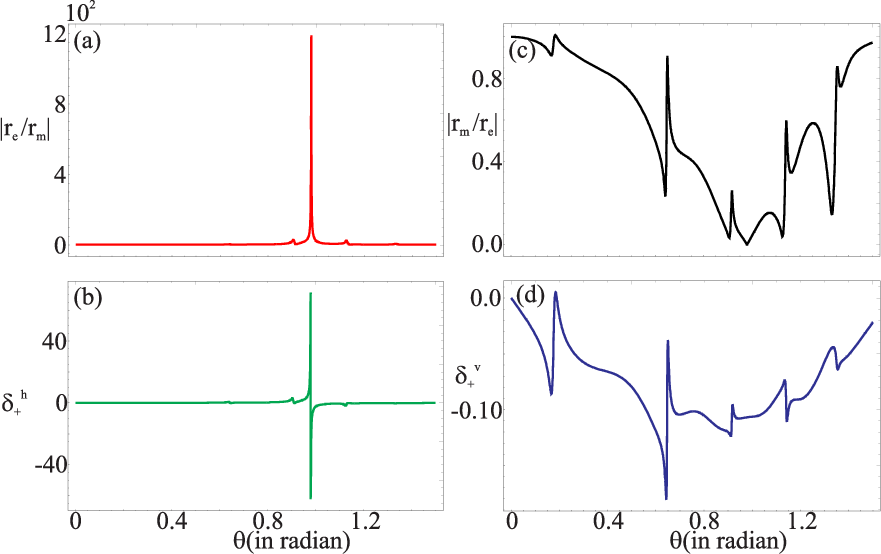}[b]
	\caption{(a) The ratio $|r_e/r_m|$ versus $\theta$ (b) The horizontal PSHE versus $\theta$ (c) The ratio $|r_m/r_e|$ versus $\theta$ (d) The vertical PSHE versus $\theta$. The parameters are $\gamma_{bl}=1.36$meV, $\gamma_{bd}=0.68$meV, $\gamma_{cl}=1.36$meV, $\gamma_{cd}=0.8$meV, $\gamma_{dl}=0.8$meV, $\gamma_{dd}=0.5$meV, $\beta=0.0184$, $g=-1$, $f=1$, $\Delta=2$meV, $\epsilon_1=\epsilon_2=2.22$, $d_1=0.2\mu$ m, $d_2=5\mu$ m, and $\Omega_c=0$}
	\label{figure2}
\end{figure}
\begin{figure}
	\centering
	\includegraphics[width=3.5in]{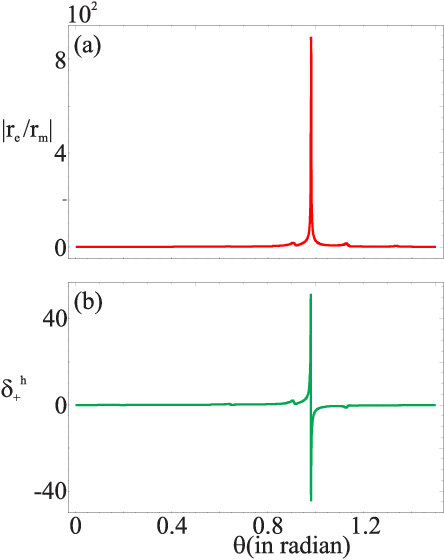}[b]
	\caption{(a) The ratio $|r_e/r_m|$ versus $\theta$ (b) The horizontal PSHE versus $\theta$. The parameters are $\gamma_{bl}=1.36$meV, $\gamma_{bd}=0.68$meV, $\gamma_{cl}=1.36$meV, $\gamma_{cd}=0.8$meV, $\gamma_{dl}=0.8$meV, $\gamma_{dd}=0.5$meV, $\beta=0.0184$, $g=-1$, $f=1$, $\Delta=2$meV, $\epsilon_1=\epsilon_2=2.22$, $d_1=0.2\mu$ m, $d_2=5\mu$ m, and $\Omega_c=6$meV}
	\label{figure3}
\end{figure}

\section{results and discussion}

We conduct a comprehensive investigation of the horizontal and vertical polarized PSHE in our system by first considering the case when the control field is absent, i.e., $\Omega_c=0$. Under this condition, the proposed four-level system effectively reduces to a three-level configuration consisting of the states $|a\rangle$, $|b\rangle$, and $|c\rangle$. Consequently, by setting $\Omega_c=0$, the susceptibility given in Eq. (\ref{susc}) simplifies to the form $\chi = i\beta A_1/A_2$. Under this condition, the tunneling-induced interference parameter $\alpha$ plays a crucial role in simplifying the susceptibility for a fixed energy splitting $\Delta$. Specifically, for small decay rates $\gamma_{bd}=0.8$ meV and $\gamma_{cd}=0.8$ meV, the parameter $p$ attains a value of $0.83$, which corresponds to nearly perfect interference.
To quantitatively characterize the PSHE, we employ the analytical expressions given in Eqs. (\ref{h-pshe}) and (\ref{v-pshe}). In these formulations, the horizontal and vertical transverse shifts, $\delta_+^h$ and $\delta_+^v$, depend on the ratios $|r_e/r_m|$ and $|r_m/r_e|$, respectively. A direct implication of this dependence is that the horizontal and vertical polarized PSHE become significantly enhanced when $|r_e/r_m| > 1$ and $|r_m/r_e| > 1$, respectively.
To illustrate this behavior, we present the spectra of $|r_e/r_m|$ and $|r_m/r_e|$ as functions of the incident angle $\theta$, shown in Figs. \ref{figure2}(a, c). The spectrum in Fig. \ref{figure2}(a) exhibits a pronounced resonance near $\theta = 0.98$ rad, where the ratio $|r_e/r_m|$ reaches a maximum value of approximately 1200. This extreme enhancement originates from the near-complete suppression of the magnetic reflection coefficient, while the electric reflection coefficient remains finite. The pronounced asymmetry between the electric and magnetic responses amplifies $|r_e/r_m|$, thereby giving rise to a substantial enhancement of the horizontal PSHE under these specific conditions. In contrast, the ratio $|r_m/r_e|$ shows no significant enhancement for the same parameters, as depicted in Fig. \ref{figure2}(c).
Next, we turn to the horizontal and vertical polarized transverse shifts induced by the PSHE, focusing on the case of right-circular polarization, i.e., $\delta_+^h$ and $\delta_+^v$. Due to the intrinsic symmetry of the PSHE, the corresponding shifts for left-circular polarization, $\delta_-^h$ and $\delta_-^v$, exhibit the same magnitude but occur in the opposite direction. Following the above conditions, we plot $\delta_+^h$ and $\delta_+^v$ as functions of $\theta$, shown in Figs. \ref{figure2}(b, d). A pronounced horizontal transverse shift emerges around $\theta \approx 0.98$ rad, coinciding precisely with the parameter regime where $|r_e/r_m|$ attains its maximum. This one-to-one correspondence confirms that a large PSHE is inherently linked to a strong disparity between the electric and magnetic reflection coefficients, thereby establishing $|r_e/r_m|$ as a key control parameter for spin–orbit–induced beam shifts in cavity-based systems.
Numerically, we find horizontal transverse shifts of $\delta_+^h = 71\lambda \approx 0.13$ mm and $\delta_+^h = -62\lambda \approx -0.11$ mm, both of which constitute giant beam displacements. Similarly, the vertical transverse shift reaches $\delta_+^v = -0.18\lambda$ near $\theta = 0.64$ rad.

\begin{figure}
	\centering
	\includegraphics[width=3.5in]{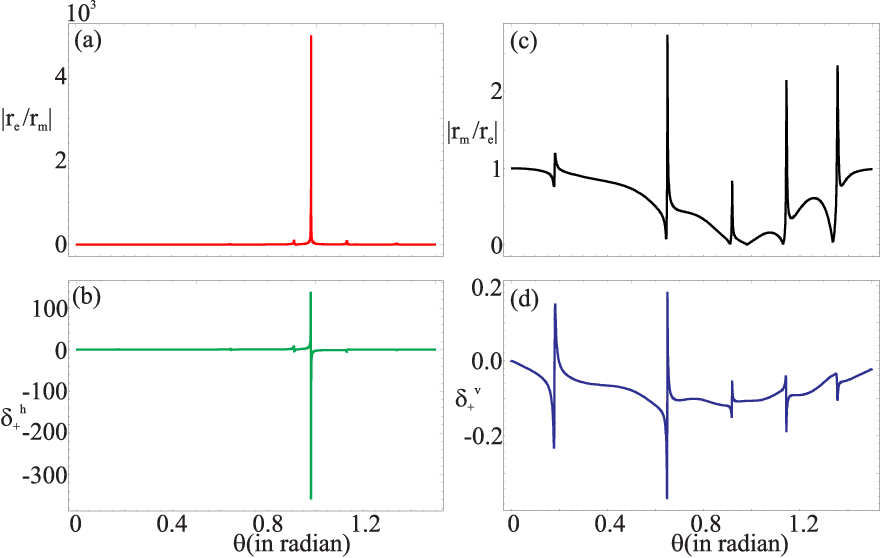}[b]
	\caption{(a) The ratio $|r_e/r_m|$ versus $\theta$ (b) The horizontal PSHE versus $\theta$ (c) The ratio $|r_m/r_e|$ versus $\theta$ (d) The vertical PSHE versus $\theta$. The parameters are $\Delta=8$meV, $\gamma_{bd}=1.36$meV, $\gamma_{cd}=1.6$meV and $\Omega_c=0$, the remaining parameters remains the same as shown in Fig. Fig. \ref{figure2}.}
	\label{figure4}
\end{figure}
\begin{figure}
	\centering
	\includegraphics[width=3.5in]{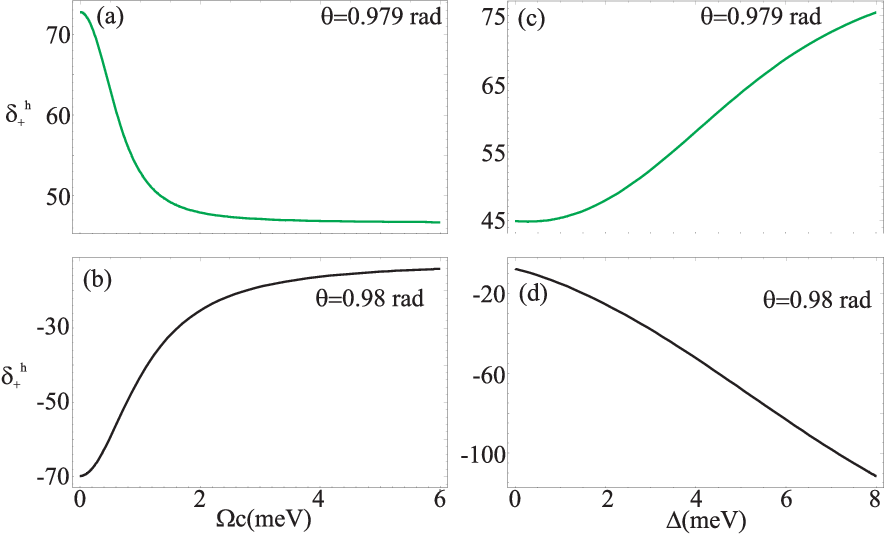}[b]
	\caption{(a) The horizontal transverse shift $\delta_+h$ versus $\Omega_c$ when $\Delta=2$meV and $\theta=0.979$ (b)  The horizontal transverse shift $\delta_+h$ versus $\Omega_c$ when $\Delta=2$meV and $\theta=0.98$. (c)  The horizontal transverse shift $\delta_+h$ versus $\Delta$ when $\Omega_c=2$meV and $\theta=0.979$ (d)  The horizontal transverse shift $\delta_+h$ versus $\Delta$ when $\Omega_c=2$meV and $\theta=0.98$.The values for the remaining parameters are identical to those used in Fig. \ref{figure2}.}
	\label{figure5}
\end{figure}
\begin{figure}
	\centering
	\includegraphics[width=3.5in]{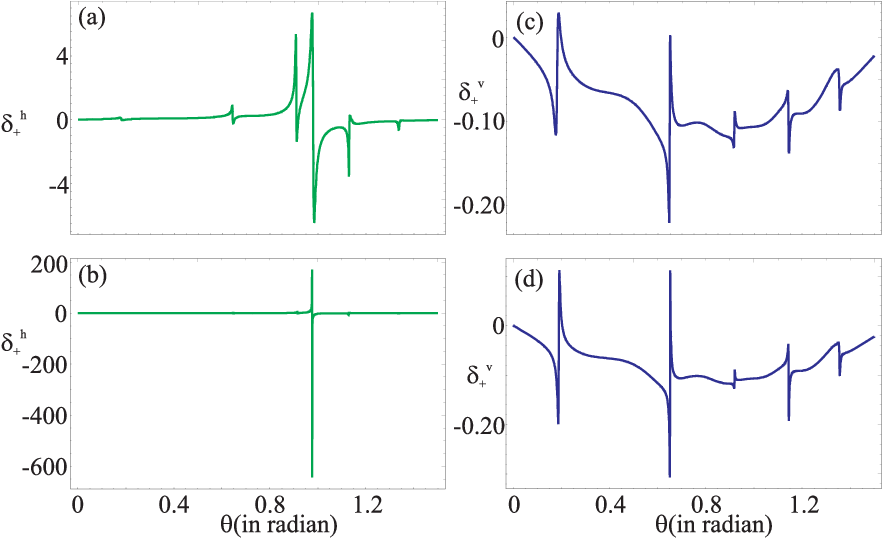}[b]
	\caption{(a, c) The horizontal transverse shift $\delta_+^h$ versus $\theta$ by considering $\epsilon_1=\epsilon_2=2.22+0.04 i$ (b, d)The horizontal transverse shift $\delta_+^h$ versus $\theta$ by considering $\epsilon_1=2.22+0.04i$ and $\epsilon_1=2.22-0.04i$.The values for the remaining parameters are identical to those used in Fig. \ref{figure4}.}
	\label{figure6}
\end{figure}

Conversely, when the control beam is switched on ($\Omega_c \neq 0$), the coupling transitions $|b\rangle \leftrightarrow |d\rangle$ and $|c\rangle \leftrightarrow |d\rangle$ become activated, which significantly enhance the probe absorption in the quantum well medium through the quenching of tunneling-induced interference. Under this condition, we plot the ratio $r_e/r_m$ and the horizontal transverse shift $\delta_+^h$ as functions of the incident angle $\theta$, as shown in Fig. \ref{figure3}. In contrast to the case of $\Omega_c = 0$, both the spectral profile of $r_e/r_m$ and the magnitude of $\delta_+^h$ are notably reduced. This reduction highlights the crucial role of the control field in tuning the balance between electric and magnetic reflection responses, thereby offering an additional degree of control over the PSHE in the quantum well system.

As previously reported, the strength of tunneling-induced interference diminishes with an increase in temperature up to 30 K \cite{H-C-Liu-tra}, which in turn enhances the dephasing rates, such as $\gamma_{bd}=1.36$ meV and $\gamma_{cd}=1.6$ meV. Importantly, the coupling strength of resonant tunneling, $\Delta$, plays a fundamental role in tailoring this interference and thus provides an effective means to manipulate the PSHE. By fixing the dephasing parameters at $\gamma_{bd}=1.36$ meV and $\gamma_{cd}=1.6$ meV, and considering $\Delta=8$ meV with $\Omega_c=0$, we evaluate the ratios $|r_e/r_m|$, $|r_m/r_e|$, along with the corresponding transverse shifts $\delta_+^h$ and $\delta_+^v$ as a function of the incident angle $\theta$, as shown in Fig. \ref{figure4}. Remarkably, under the influence of a large tunneling strength $\Delta$, the ratio $|r_e/r_m|$ attains a giant value of 4828 near $\theta=0.98$ rad. This corresponds to an extraordinary horizontal shift, yielding $\delta_+^h = 135 \lambda \approx 0.25$ mm and $\delta_+^h = -335 \lambda \approx -0.63$ mm. Such results signify the realization of giant horizontal transverse shifts enabled by strong resonant tunneling. In contrast, the vertical shift exhibits only a marginal enhancement at different incident angles, as illustrated in Fig. \ref{figure4}(d).

To further clarify the interplay between the control field strength $\Omega_c$ and the resonant tunneling coupling $\Delta$, we analyze the horizontal-polarized PSHE as a function of these parameters, as illustrated in Fig. \ref{figure5}. For $\theta=0.979$ rad, the dependence of the transverse shift $\delta_+^h$ on $\Omega_c$ is presented in Fig. \ref{figure5}(a). The spectrum reveals that the positive transverse shift is initially large and gradually decreases with increasing $\Omega_c$. In contrast, when the incident angle is slightly varied to $\theta=0.98$ rad, the spectrum shown in Fig. \ref{figure5}(b) exhibits a negative transverse shift. In this case, the magnitude of the shift is pronounced at small values of $\Omega_c$ and decreases progressively as $\Omega_c$ increases.
A complementary picture emerges when we fix $\Omega_c$ and examine the dependence of $\delta_+^h$ on $\Delta$ for the two incident angles $\theta=0.979$ rad and $\theta=0.98$ rad, as shown in Figs. \ref{figure5}(c, d). Notably, both the positive and negative transverse shifts begin with small values at low tunneling strengths and grow monotonically with increasing $\Delta$. These results highlight the tunability of the PSHE through the delicate balance between the control field and resonant tunneling, where even small variations in $\theta$ dictate whether giant positive or negative transverse shifts can be realized.

 In the preceding analysis, the permittivity of the cavity slabs was assumed to be absorption-less, enabling us to examine the intrinsic behavior of the PSHE. A natural question then arises: how does the PSHE respond when the cavity slabs are absorptive or possess gain? To address this, we extend our study by incorporating absorption and gain into the cavity design.
 First, we consider the case where both cavity slabs exhibit weak absorption, specifically by setting $\epsilon_1=\epsilon_2=2.22+0.04i$. The corresponding horizontal and vertical transverse shifts, $\delta_+^h$ and $\delta_+^v$, are evaluated as functions of the incident angle $\theta$, as shown in Figs. \ref{figure6}(a, c). The results reveal a noticeable reduction in both horizontal and vertical PSHE compared with the lossless scenario, indicating that absorption suppresses the spin-dependent beam shifts by damping the interference pathways responsible for their enhancement.
 The situation becomes particularly fascinating when the system is engineered such that one cavity slab is absorptive while the other is gain-assisted. For this configuration, we select $\epsilon_1=2.22+0.04i$ and $\epsilon_2=2.22-0.04i$, and compute the transverse shifts $\delta_+^h$ and $\delta_+^v$, as illustrated in Figs. \ref{figure6}(b, d). Remarkably, under this balanced gain–loss condition, the horizontal transverse shift exhibits giant enhancements, reaching values as high as $\delta_+^h=200\lambda \approx 0.38$ mm and $\delta_+^h=-600\lambda \approx -1.14$ mm, as shown in Fig. \ref{figure6}(b). These extraordinary shifts stem from the non-Hermitian interplay between gain and loss, which induces a resonant amplification of the spin-orbit coupling within the cavity. In contrast, the vertical shift shows only a modest increase, as depicted in Fig. \ref{figure6}(d), highlighting the anisotropic sensitivity of the PSHE to gain–loss engineering.
 Thus, the incorporation of balanced absorption and gain in cavity slabs not only enriches the fundamental physics of the PSHE but also provides a practical route toward achieving giant and controllable spin-dependent beam shifts for photonic applications.

\section{summary}
     
In summary, we have proposed and analyzed a scheme for actively controlling the PSHE in the reflected beam of a cavity system incorporating an asymmetric double quantum well medium. By employing the transfer matrix method, we systematically evaluated the reflection coefficients for both TE- and TM-polarized probe beams, thereby capturing the spin-dependent transverse shifts. Our study demonstrates that the optical characteristics of the quantum well, modulated through an external control beam and resonant tunneling, give rise to a giant PSHE. Moreover, the incorporation of loss- and gain-assisted cavity slabs further amplifies the horizontal PSHE, leading to a more pronounced spin-dependent response. These findings not only deepen the understanding of PSHE in engineered semiconductor structures but also highlight new possibilities for tailoring spin–orbit light interactions in photonic devices.



\end{document}